\documentclass[number]{elsarticle}

\usepackage[utf8]{inputenc}
\usepackage[T1]{fontenc}
\usepackage[english]{babel}

\usepackage{listings}
\usepackage{lstlang3}
\usepackage{url}
\usepackage{amsmath}

\usepackage{amssymb}
\usepackage{mathpartir}
\usepackage{xcolor}
\usepackage{multirow}
\usepackage{alltt}
\usepackage{url}
\usepackage{xspace}
\usepackage{enumerate}
\usepackage{bbm}
\usepackage{stmaryrd}
\usepackage{url}
\usepackage{tabularx}
\usepackage{listings}
\usepackage[font=footnotesize,caption=false]{subfig}
\usepackage{floatflt}

\usepackage{tikz}
\usetikzlibrary{petri}
\usetikzlibrary{positioning}
\usetikzlibrary{arrows}
\usepgflibrary{shapes}
\usetikzlibrary{decorations.pathreplacing}
\usetikzlibrary{calc}
 
\newif\iflong

\newcolumntype{C}{>{${}}c<{{}$}}
\newcolumntype{L}{>{${}}l<{{}$}}
\newcolumntype{R}{>{${}}r<{{}$}}

\tikzstyle{transition} =[very thick, rectangle, draw, inner xsep=2mm, inner ysep=0.75mm]
\tikzstyle{vtransition}=[very thick, rectangle, draw, inner ysep=2mm, inner xsep=0.75mm]

\tikzstyle{place}=[circle, draw, minimum size=4ex]
\tikzstyle{every label}=[font=\sf\footnotesize]

\tikzstyle{pre}+=[>=stealth]
\tikzstyle{post}+=[>=stealth]
\tikzstyle{readarc}=[pre, >=*, shorten <=0pt]
\tikzstyle{prio}=[draw, ->, orange, >=stealth, shorten >=1.25pt, shorten <=1.25pt, densely dashed]
\tikzstyle{poids}=[font=\scriptsize\sf]

\tikzstyle{action}=[rectangle, draw, color=black!80, thin, dotted, inner xsep=0.4ex, inner ysep=0.1ex]

\tikzstyle{gil common}=[color=black, thin, draw=none, fill=none, dash pattern=]

\tikzstyle{gil search}=[gil common, draw, densely dashed, |->, >=stealth]
\tikzstyle{gil strong search}=[gil search, draw, |->>]
\tikzstyle{gil search syntax}=[gil common, draw, ->, >=stealth]
\tikzstyle{gil context}=[gil common, draw, {[-)}]

\tikzstyle{gil close context}=[gil common, draw, {[-)}]
\tikzstyle{gil open context}=[gil common, draw, {]-)}]
\tikzstyle{gil dotted context}=[gil common, draw, loosely dotted,|-|]
\tikzstyle{gil open search}=[gil search, draw, {|->}]
\tikzstyle{gil close search}=[gil search, draw, {|->}]
\tikzstyle{gil time context}=[gil common, draw, [-)]
\tikzstyle{gil open strong search}=[gil search, draw, {|->>}]
\tikzstyle{gil close strong search}=[gil search, draw, {|->>}]

\tikzstyle{gil segment}=[gil common, draw, [-)]
\tikzstyle{gil exists}=[gil common, draw,kite,midway,kite vertex angles=60, inner sep=0.4ex]
\tikzstyle{gil always}=[gil common, draw, gil exists, rectangle, inner ysep=0.8ex]
\tikzstyle{gil boite}=[gil common, draw, dotted, thick, rounded corners=4pt]
\tikzstyle{delta}=[gil common, draw,isosceles triangle, anchor=apex, rotate=90, inner sep=0.3ex]

\tikzstyle{gil interval}=[gil common, draw, decorate,decoration={brace,mirror}]

\tikzstyle{every pin}=[gil common, pin distance=0.4ex]
\tikzstyle{every pin edge}=[gil common, draw,-]

\tikzstyle{adroite}=[gil common, anchor=east, inner sep=0em, xshift=2ex]
\tikzstyle{audessus}=[gil common, anchor=south east, inner sep=0em, yshift=1.2ex]
\tikzstyle{audessous}=[gil common, anchor=north east, inner sep=0em, yshift=-1.2ex]
\tikzstyle{inlaudessous}=[gil common, anchor=east, inner sep=1ex, yshift=-0.5ex]
\tikzstyle{inlaudessous2}=[gil common, anchor=east, inner sep=2.2ex, yshift=-0.5ex]

\tikzstyle{timint}=[gil common, midway,above=-0.5ex]

\tikzstyle{snippet}=[gil common, draw, rectangle, rounded corners=0pt, inner ysep=0.15ex, inner xsep=1ex, semithick,
                     densely dotted, draw=black, text width=]

\newcommand{\pop}[1]{\ensuremath{\mathop{\text{\textsf{\textbf{{#1}\,}}}}}}

\begin{document}

\begin{frontmatter}

\title{Real-Time Model Checking Support for {{AADL}}}

\author[laas,ut]{B.~Berthomieu}
\author[irit,ut]{J.-P.~Bodeveix}
\author[laas,ut]{S.~{Dal~Zilio}\corref{cor1}}
\ead{dalzilio@laas.fr}
\author[irit,ut]{M.~Filali}
\author[laas,ut]{D.~{Le~Botlan}}
\author[irit,ut]{G.~Verdier}
\author[laas,ut]{F.~Vernadat}


\cortext[cor1]{Corresponding author}
\address[irit]{CNRS, IRIT, 118 route de Narbonne,
  F-31062 Toulouse, France}
\address[laas]{CNRS, LAAS, 7 avenue du colonel Roche, F-31400
  Toulouse, France}
\address[ut]{Univ de Toulouse, F-31400 Toulouse, France}


\begin{abstract}
  We describe a model-checking toolchain for the behavioral
  verification of AADL models that takes into account the realtime
  semantics of the language and that is compatible with the AADL
  Behavioral Annex. We give a high-level view of the tools and
  transformations involved in the verification process and focus on
  the support offered by our framework for checking user-defined
  properties. We also describe the experimental results obtained on a
  significant avionic demonstrator, that models a network protocol in
  charge of data communications between an airplane and ground
  stations.
\end{abstract}


\end{frontmatter}


\section{Introduction}

The increasing complexity of the software and hardware components used
in safety critical systems has encouraged the adoption of new
architectures and computing modules, more powerful, but also more
complex than their ancestors. While these new architectures make
development and maintenance easier, it also make it more difficult to
fully understand, analyze and test these systems.

{Formal verification methods}, such as model-checking, are advocated
as one of the solutions to this consistent increase in design
complexity. While verification activities should be performed at all
stages of the development process, there are strong incentives for
carrying out as much verification as possible during the early phases,
especially during the functional and architectural design phases. To
support this trend, a number of high level system modeling languages
have been proposed---often referred to as \emph{Architecture
  Description Languages}, or simply ADL---that make it possible to
analyze a system right from the design phase.

In this paper, we describe a model-checking toolchain for the
behavioral verification of the \emph{Architecture Analysis and Design
  Language} (AADL), an ADL standardized by the SAE that can describe
both the hardware and software components of a system. The AADL
standard address the problem of specifying and analyzing
safety-critical, realtime embedded systems and is designed to support
a Model-Driven Engineering approach. A key extension to this standard
is the addition of a Behavioral Annex that refines the description of
AADL threads behavior and that can therefore be used to describe more
precisely the dynamic architecture of a system.

An advantage of AADL, compared to many other ADL, is to be based on a
precise, unambiguous semantics. Indeed, the AADL standard describe
precisely the behavior of all its components, such as: when can
messages be exchanged; how do periodic and sporadic threads interact;
how threads interact with communication or memory resources, such as
registers or communication buses; \dots Another motivation for
choosing AADL is the fact that it relies on classical hypothesis taken
when building realtime systems for its runtime; that is, AADL favors
implementability over expressiveness. This is an interesting
characteristic, since it means that every feature of the language can
be defined without resorting to any ``unrealistic'' primitives (like,
e.g., the need for a global consensus primitive). Those
characteristics are very helpful for developing semantics related
tools, like automatic code generators, schedulability analysis or
formal verification tools.

Our model-checking toolchain is based on a transformational approach,
that is, on the interpretation (the translation) of an AADL model into
a formal specification language that will take into account the
behavior of the model but also the dynamic semantics related to the
AADL standard.
We give a high-level view of the tools and transformations involved in
the verification process and focus on the support offered by our
framework for checking user-defined properties. We also report on some
initial experiments carried out in order to evaluate our framework and
give the first experimental results obtained on significant avionic
demonstrator that models a network protocol in charge of data
communications between an airplane and ground stations.

\begin{figure}[htbp]
\begin{center}
  \usetikzlibrary{shapes,arrows,positioning,fit}
  \tikzstyle{block} = [rectangle, draw, 
  text width=5em, text centered, 
  rounded corners, minimum height=3em]
  \tikzstyle{sblock} = [rectangle, draw, dashed,
  fill=gray!10,
  text width=5em, 
  text centered, 
  rounded corners, minimum height=3em]
  \tikzstyle{cloud} = [minimum height=3em, text badly ragged]
  \tikzstyle{line} = [draw, -latex]
  \tikzstyle{title} = [ minimum height=2em]
    
  \resizebox{0.89\textwidth}{!}{
  \begin{tikzpicture}[node distance = 2cm]
    \node (topcased)      [title] 	{\textbf{Topcased} (\url{www.topcased.org})};
    \node (editor) [cloud, below=0.5em of topcased]  {(semantic) editors}; 
    \node (AADL) [block, right=of editor]   {AADL}; 
    \node (ba)		[rectangle, draw,  rounded corners, minimum height=3em, right=1em of AADL, outer xsep=0cm]		{$+$BA};	
    \node (aadlba)[draw, rounded corners, fit=(AADL) (ba)]	{};
    \node (ml) 		[cloud, right=4cm of aadlba]	{modeling languages};		
    \node (fiacre) [block, below=of aadlba, text width=] 	{\begin{tabular}{c} { }\\[-0.5em] RT-Fiacre libraries\\[0.5em] \hline\\[-0.5em] Fiacre \end{tabular}};  	
    \node (pivot)	[cloud, below=of ml]	{pivot language};	
    \node (cadp) 	[sblock, below left=of fiacre] {CADP}; 	 
    \node (tina)	[block, below right=of fiacre] {Tina};	 
    \node (verif)  [cloud, below=of pivot] {verification engine}; 
    %
    \path[->] (aadlba) edge node [pos=0.5, label=right:{AADL2Fiacre}] {}  (fiacre);
    \path[->] (fiacre)  edge node [pos=0.5, label={[label distance=1em]right:{frac}}] {} (tina);
    \path[->] (fiacre) edge  node [pos=0.5, label={[label distance=1em]left:{flac}}] {} (cadp);
    \path[->] (editor) edge node [pos=0.5, label=above:{Adele}] {} (aadlba);
  \end{tikzpicture}}
  \caption{AADL to Tina toolchain}
  \label{tinaarch-fig}
\end{center}
\end{figure}
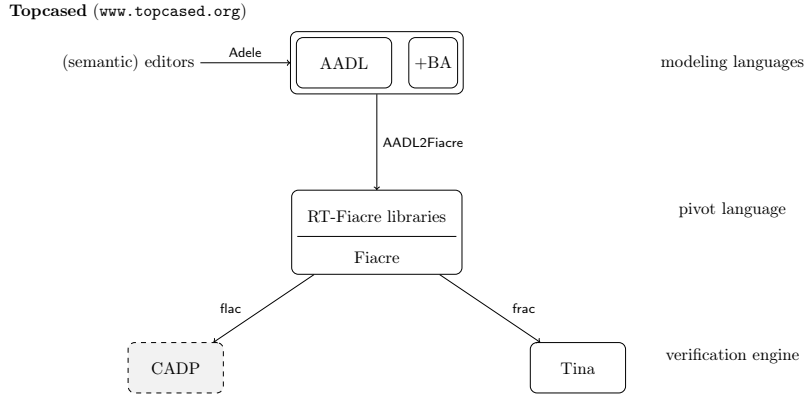


Our toolchain (see Fig.~\ref{tinaarch-fig}) is connected for its input
to Adele~\cite{ADELE}, a semantic editor for the elaboration of AADL
models. At the other end, verification activities ultimately relies on
the Tina toolset~\cite{tina}, that provides state-space generation and
model-checking algorithms for timed Petri Nets. In-between, the
generation of Tina models from an AADL description relies on the use
of an intermediate formal specification language, named
Fiacre~\cite{filfmvte2008}. Fiacre offers a formal framework to
express and inspect the behavioral and timing aspects of the
system. The intermediate Fiacre model provides a formal representation
of a system behavior that is suitable for analysis using a
model-checking tool. Actually, most of the same toolchain can be used
to derive formal specifications for the Tina and the CADP
model-checker~\cite{cadp}.

The transformation from AADL to Fiacre is based on a Model Driven
Engineering approach---where the adaptation and integration between
tools is ensured by model-based techniques---and has been integrated
into an Eclipse-based toolkit for system engineering called
Topcased~\cite{ttptosfcasd2006}. Topcased provides an open source,
model oriented set of tooling and standard implementations and AADL
was among the first languages supported in this project.

Our current toolchain is the result of the refinement and maturation
of several previous versions of the AADL2Fiacre
interpretation~\cite{aadl2fcr,FVAMFT}. In this most recent iteration
of our tool, we have focused on the modularity of the transformation
with the goal to increase its maintainability and to simplify the
proof of its correctness. Indeed, our previous implementation were
based on a monolithic interpretation, that is supposed to generate
fewer states but that was more delicate to debug and extend. One of
the results obtained from our experimentations is that it is possible
to follow a compositional approach for the encoding without degrading
the performances; actually, we observe that following a more
compositional approach makes it is easier to take benefit from
symmetries in the system and to recover static dependencies than can
help reduce the number of interleaving in the generated state space.

\paragraph*{Outline:}
We briefly describe the AADL execution model in
Sect.~\ref{sec:aadl-execution-model} and focus on the behavior of
threads and their interactions with communication events.  Next, we
give a high-level view of the tools and languages involved and
illustrate the successive transformations required by our verification
process. We describe the Fiacre language and its support for checking
user-defined, realtime properties.  In particular, we show how to use
realtime specification patterns to check properties on the
interpretation of an AADL model. Before concluding, we describe in
Sect.~\ref{sec:experiments} the results obtained on an AADL
demonstrator.





\section{AADL Execution Model and the Behavioral Annex}
\label{sec:aadl-execution-model}

The AADL standard has been designed with the goal to provide a precise
description of both the software components of a system (such as
processes, threads, data, \dots) as well as the execution platforms
supporting them (processors, devices, buses, memory, \dots). The
language has both a graphical and a textual syntax and includes all
the usual concepts found in a component-based languages: components
are typed and are described using a semi-structured set of properties;
the interface of a component can be defined using the notion of
features; connections between components can be described using a
notion of links.

The AADL execution model is suitable to describe real-time systems
because it includes the main types of dispatch protocols for threads
(periodic, aperiodic, sporadic, background) and the standard
scheduling properties (period, priority, deadline, WCET, scheduling
policy, \dots). The language also includes the basic methods for
interaction; components can communicate through ports, synchronous
calls, and shared data. The AADL notion of process is the unit for
describing the dynamic semantics of a system. A process represents a
virtual address space, or a partition, that includes a program and all
its sub-components. A process must contain at least one thread (or
thread group) that represents a sequential flow of execution. Threads
are the only AADL components that can be scheduled. The AADL
Behavioral Annex is used to add specific real-time properties to each
component of the dynamic design model and to define the software
behavior at the thread level.  We can define the real time properties
of threads by setting specific properties in the AADL specification,
like for instance the dispatch protocol (periodic or sporadic), the
period (time) and the deadline (time). An example of thread
declaration using the behavioral annex can be seen in the AADL code
snippet of Listing~\ref{apota}. (An example of AADL graphical diagrams
is given in
page~\pageref{graph-atc-fig}.)\\

\begin{lstlisting}[%float,%
xleftmargin=0.5em,xrightmargin=0.5em,%
keywordstyle=\color{black}\bfseries,
numbers=left, numberstyle=\tiny,%
basicstyle=\scriptsize,frame=single,caption=Example of AADL behavior
description, label=apota,language=aadl]
THREAD thApplis
FEATURES
{...}
END thApplis;

THREAD IMPLEMENTATION thApplis.others
SUBCOMPONENTS
  app_msg : DATA types::msg.impl;
{...}
PROPERTIES
  Dispatch_Protocol => Periodic; Deadline => 10 ms; Period => 10 ms;
{...}
ANNEX behavior_specification {**
  states  -- States Declaration
   start : initial state;  pending : complete state; 
   register : complete state; disreg : complete state
   ...
 transitions
   start   -[ ]-> pending { app_msg.req := 0; app_msg.dat := 0;  };
   pending -[ on app_msg.req = Reg ]-> register { app_msg.req := 0; };
   pending -[ on app_msg.req = Disreg ]-> disreg { app_msg.req := 0; };
{...}
**};
END thApplis.others;
\end{lstlisting}

AADL is supported by several tools like the OSATE initial framework,
which has been integrated into the Topcased environment and extended
with OSATE-BA, the behavioral annex syntax analyzer. For editing
models, Adele is a graphical editor which permits to create
(graphical) AADL diagrams in Topcased and to generate AADL source
code. Beside this set of tools for the generation and lexical analysis
of AADL models, we describe a methodology and a set of tools for the
formal verification of AADL specifications.  For behavioral
verification, we can only focus on a subset of AADL (In particular we
do not take into account hardware components). We briefly describe the
semantics of threads, their scheduling, and the communication through
ports and shared data. Modes are not modeled yet, but we plan to
integrate them in our tool.


\paragraph*{{Communication through ports}} Communication, and the way
it interacts with the scheduling of processes, is an important part of
the AADL standard. AADL provides three types of ports---data, event
and event data ports---that can be used to transmit data and control
and describe the interface of a component.

Data transmitted through ports is typed. Each input port is associated
with a fresh variable that describes the state of the port. If a port
has received nothing between two thread dispatches this variable is
set to false. Each event or event data input port is also associated
with a buffer that stores the data---or the number of events---sent
through connected output ports. On thread dispatch, these inputs
buffers are copied into the local memory of the thread. Properties can
be used to customize the behavior of event and event data ports. For
instance, the property \verb+Queue_size+ determines the maximum number
of events or event data that can be received, while
\verb+Overflow_handling_protocol+ describes the behavior of the port
in case of overflow. (There are two default policies for overflow,
drop newest and drop oldest.) The use of the \verb+Queue_size+
property is useful to generate a finite-state system from a model.

\begin{figure}[!h]
\begin{center}
\includegraphics[width=8cm]{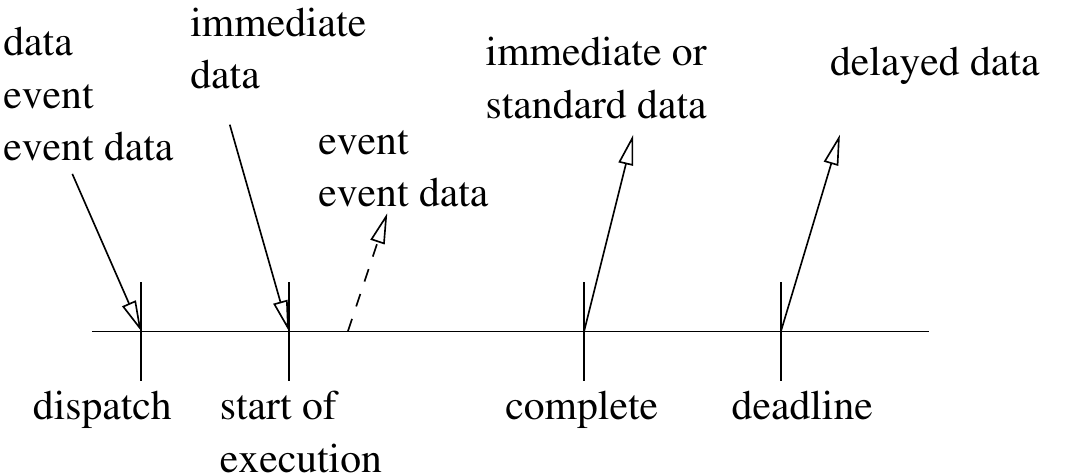}
\end{center}
\caption{\label{aadlcom}\emph{communication through ports in AADL.}}
\end{figure}

The diagram in Fig.~\ref{aadlcom} depicts the typical interaction
between data communication through ports and thread dispatching. The
axis on this diagram list the four possible state of a periodic
thread: \verb+dispatch+ (the scheduler allows the thread to run);
\verb+start+; \verb+complete+ (the thread starts, respectively en, its
computation); and \verb+deadline+ (that should always occur after a
\verb+complete+ event, if the system is schedulable).

Data ports have the simplest behavior: data is sent at the end of the
thread execution, or at deadline, and is received at the next dispatch
of the receiving thread. At the opposite, event and event data ports
can send an event (resp. an event data) anytime during the execution
of a thread. Events and event data are queued in the destinations
ports. Input event and event data ports are delivered at the dispatch
of the thread. Data communications between periodic threads can be
declared as immediate or delayed. If the connection is delayed, data
is sent at the deadline of the sending thread. If the connection is
immediate, the receiving thread must wait the sending thread to
complete. The received data will be available at the start of its
(next) execution. All the possible combination of communication
behaviors have been taken into account in our formal interpretation of
AADL.


\paragraph*{Communication through shared variables} As with all AADL
components, data has a type and an implementation.  The internal
structure of the data is described in the data implementation. It is
possible to specify whether different components have a shared access
to a data subcomponent using the \verb+require_data_access+
connector. Correspondingly, the \verb+provide_data_access+ connector
is used to state that a component allows other components access to
one of its data subcomponent. The concurrency protocol used to access
a data is defined by a data property called
\verb+concurrency_control_protocol+. This concurrency protocol can be
implemented through different concurrency control mechanisms such as
mutex, semaphore~\ldots Concurrency protocols are a significant source
of variability in the definition of the AADL syntax.  We take into
account this variability in our interpretation of AADL to Fiacre (and
the possibility to extend the language with new, user-defined,
protocols) by providing an extensible library of protocols and
providing supports for checking the correctness of these
protocols. That is, support to prove that the semantics of a protocol
(such as mutual exclusion) is preserved by our interpretation.



\section{The Fiacre Specification Language  and Realtime Properties}
\label{sec:fiacre-spec-lang}

Our verification toolchain is based on a transformation from AADL into
an input format suitable for our model-checking tools. This
transformation relies on the use of the Fiacre specification language
to facilitate the processing; simplify the maintainability of our tool
(e.g. when the AADL standard is revised); and simplify the reasoning
on the correctness of the transform.

The Fiacre language has been designed in the context of the Topcased
project~\cite{ttptosfcasd2006} to serve as an intermediate format
between high-level description languages and formal verification
tools. The use of a formal intermediate modeling language has several
benefits. First, it helps reduce the semantic gap between high-level
models and the input format of verification tools that often relies on
low level formalisms, such as Petri Nets or process algebra. Second,
the use of a {formal language} makes it possible to define precisely
the semantics of the input language ``only once'' and to share this
work among different verification toolchains. This is particularly
helpful when we try to address emergent system modeling language,
whose semantic evolves rapidly.

\subsection{An Example of Fiacre Specification: the Periodic Thread Controller}

Fiacre is a formal specification language designed to represent both
the behavioral and timing aspects of real-time systems. Fiacre
supports two of the most common communication paradigms: communication
through shared variable and synchronization through (synchronous)
communication ports. In the latter case, it is possible to associate
time and priority constraints to communication over ports. The design
of Fiacre is inspired from decades of research on concurrency theory
and real-time systems theory. For instance, its timing primitives are
borrowed from Time Petri nets, while the integration of time
constraints and priorities into the language can be traced to the BIP
framework~\cite{bipaadl16}.  For composing components, Fiacre
incorporates a parallel composition operator and a notion of gate
typing which were previously adopted in Lotos-NT. We briefly describe
the language. The detailed syntax and formal semantics of the Fiacre
can be found in~\cite{filfmvte2008}.

Fiacre programs are stratified in two main notions: processes and
components. {Processes} describes the behavior of sequential
components. A process is defined by a set of control states, each
associated with an expression that specifies state transitions
(introduced by the keyword \verb+from+). Expressions are built from
deterministic constructs available in classical programming languages
(assignments, conditionals, sequential composition, ...);
non-deterministic constructs (choice and non-deterministic
assignments); communication events on ports; and jump to next state
(introduced by the keywords \verb+loop+ and \verb+to+). {Components}
describes the composition of processes, possibly in a hierarchical
manner. A component is defined as a parallel composition of components
and processes communicating through ports and shared variables. A
component can be used to restrict the access mode and visibility of
shared variables and ports, to associate timing constraints with
communication ports and to define priority between communication
events. We give an example of Fiacre specification in
Listing~\ref{periodic}.

\begin{lstlisting}[float,%
numbers=left, numberstyle=\tiny,%
xleftmargin=0.5em,xrightmargin=0.5em,%
keywordstyle=\color{black}\bfseries,
basicstyle=\scriptsize,frame=single,caption=Example of Fiacre process
(interpretation of AADL periodic threads),
label=periodic,language=FIACRE]
type p_state is union p_idle | p_rdy | p_err end

process periodic[d: none, c: none, dl: none, w: none] is
  states s0, sched_error
  var st: p_state := p_rdy
  init to s0
  from s0 
    select
       c ; st := p_idle ; loop
    [] (on st=p_rdy) ; dl ; loop
    [] d ; loop
    [] w ; st := (st=p_rdy ? p_err : p_rdy) ; to s0
    unless (on st=p_err) ; wait [0,0] ; to sched_error
    end

component main is
  port w : none in [20,20], d : none in [0,0] , ...
  priority c > dl > d ,  c' > dl' > d', ...
  par periodic[d, c, dl, w] || periodic[d', c', dl', w'] || ... end

property req1 is ltl []((main/1/event dl) => <>(main/1/event d))
\end{lstlisting}

The process \verb+periodic+, defined in Listing~\ref{periodic}, models
the behavior of an AADL periodic thread. We consider the simplest
case, where the period is equal to the deadline and where no data is
exchanged (the ports have the type \verb+none+). The process may
interact with its environment through four external ports, passed as
parameters of the process declaration (line 3 of
Listing~\ref{periodic}): a port for dispatch (\verb+d+), complete
(\verb+c+) and deadline events (\verb+dl+) and a port (\verb+w+) that
is used to check that the thread has stopped executing---it is idle
\verb+(s = p_idle+)---before it reaches a new period. The
\verb+periodic+ process loops on the state \verb+s0+ and relies on a
local variable (\verb+st+) to encode the current condition of the
thread (idle, ready or error). The \verb+select+ operator is used to
model a non-deterministic choice between several transitions,
separated by the symbol \verb+[]+, whereas the keyword \verb+unless+
is used to assign the highest priority among a set of transitions.
Hence, if \verb+st+ has the value \verb+p_err+, the process
necessarily goes to the state \verb+sched_error+ where it blocks (line
13). In this transition, the \verb+wait+ operator is used to express
the fact that the change is instantaneous (it takes a duration in the
time interval $[0,0]$).

The component \verb+main+ is used to create several instances of the
periodic thread. In our encoding of periodic threads, we declare a new
port \verb+w+ for every instance of the process \verb+periodic+; this
port is associated to a temporal constraint of the form $[T;T]$, where
$T$ is the period of the thread (in our example, $T=20$). On the
opposite, the ports \verb+d+, \verb+c+ and \verb+dl+ are instantaneous
(they are associated to the time constraint $[0;0]$) and constrained
by a priority relation of the form \verb+c > dl > d+.

We can express (a very weak form of) the real-time requirements of the
periodic thread using formulas in a temporal logic, like LTL for
instance. For example, we can express the requirement that---in the
absence of scheduling errors---a deadline event is always followed by
a dispatch. This property can be easily expressed in LTL with a
formula of the form:
\begin{equation}\label{eq:req1}
  \Box\, \text{deadline} \Rightarrow \Diamond\, \text{dispatch}
\end{equation}

A strong limitation of an approach based on LTL model-checking is that
it is not possible to express timing constraints like, for example,
that the dispatch should happen before $T$ time units of the deadline
(where $T$ is the period of the thread). Another limitation is that it
is necessary to understand how events from the initial AADL model are
translated into events or states in the Fiacre model. In the following
section, we show an extension to the Fiacre language that makes it
easier to express timed temporal properties. This extension was
specially added to alleviate the two limitations that we just pointed
out.


\subsection{Expressing Real-Time Requirements in Fiacre}
\label{sec:Real-time-Rq}

The chief purpose of the Fiacre language is to express the behavior of
realtime, reactive systems. Nonetheless, it is also possible to
declare, inside a Fiacre model, a set of properties that should be
valid on the model. Each property is declared in the Fiacre model using
the keyword \verb+property+; for example, line 21 of
Listing~\ref{periodic} declares a requirement equivalent to the LTL
property~\eqref{eq:req1}.

In this section, we briefly describe the set of realtime specification
patterns available in our framework. A complete description of the
language is given in~\cite{DalzilioS:fmics2012patterns}. Our language
extends the property specification patterns of Dwyer et
al.~\cite{ppsfsv1999} with the ability to express time delays between
the occurrences of events. The result is expressive enough to define
properties like the compliance to deadline, bounds on the worst-case
execution time, etc.  The advantage of this approach is to provide a
simple formalism to non-experts for expressing properties. Another
benefit is that properties expressed with this pattern language can be
directly used with our model-checking tools. The pattern language
follows the same classification that in Dwyer's work, with patterns
arranged in categories such as occurrence or order patterns. In the
following, we study examples of \textit{response} and \textit{absence}
patterns. 



\paragraph{Response pattern with delay} This category of patterns can
be used to express delays between events, like for example constraints
on the Worst Case Execution Time of a task. The typical example of
response pattern states that every occurrence of an event, say $e_1$,
must be followed by an occurrence of an event $e_2$ within a time
interval $I$. This pattern is denoted:
\[\tag{leadsto-within} e_1 \pop{leadsto} e_2 \pop{within} I~.\]

Events that are observable at the Fiacre level are: a process entering
or leaving a state; a variable changing value; a communication through
a port. Therefore, considering the (sketch of the) interpretation of
AADL threads in Fiacre given in the previous section, we can use the
notation $\text{t/event} e$ to refer to a synchronization over the
port $e$ on the controller process for the thread t. Hence, we can
check that the execution time of the thread \verb+periodic+ is less
than $T$ units of time with the following requirement, meaning that
the time between a dispatch and a completion is always less than $T$:\\

\noindent\lstinline[
keywordstyle=\color{black}\bfseries,%
language=FIACRE]%
!property req2 is (main/1/event c) leadsto (main/1/event d) within [0; T]!



\paragraph{Absence pattern with delay} This category of patterns can
be used to specify delays within which activities must not occur. A
typical pattern in this category can be used to assert that an
activity, say $e_2$, cannot occur between $d_1$--$d_2$ units of time
after the occurrence of an activity $e_1$. This requirement
corresponds to a basic absence pattern in our language:
\[\tag{absent-after} \pop{absent}
e_2 \pop{after} e_1 \pop{within} [d_1; d_2]~.
\]
An example of use for this pattern is the requirement that we cannot
have two dispatch events for the same periodic thread in less than the
period, say $T$:\\

\noindent\lstinline[
keywordstyle=\color{black}\bfseries,%
language=FIACRE]%
!property req3 is absent (main/1/event d) after (main/1/event d) within ]0; T[!\\

\noindent A more complicated example of requirement is to impose that,
in every run such that a dispatch is followed by a completion in less
than $T$, then there are no scheduling error. This requirement can be
expressed using the composition of the properties \verb+req2+ and
\verb+req4+:

\noindent\lstinline[
keywordstyle=\color{black}\bfseries,%
language=FIACRE]%
!property req4 is absent (main/1/state sched_error)!


\subsection{Behavioral Verification with Tina}

The ``meaning'' of a Fiacre program can be expressed as a Timed
Transition System (TTS) \cite{HenzingerMP91}, defined from the states of the system
processes and from timed transitions between these states. The
\emph{frac} compiler can be used to build a TTS from a Fiacre
program. The Tina verification toolbox~ \cite{tina} offers several
tools to work with TTS files. %
For instance, for verification
purposes, TTS specifications can be used by \emph{selt}---a
model-checker for a State-Event version of Linear Temporal Logic
(LTL)---and by \emph{muse}---a model-checker for the $\mu$-calculus.

Beside the usual analysis facilities of similar environments, the
essential components of the Tina toolbox are state space abstraction
methods and model checking tools that can be used for the behavioral
verification of systems. This is in contrast with the broader notion
of functional verification, in that we attempt to use formal
techniques to prove that requirements are met, or that certain
undesired behaviors cannot occur---like for instance
deadlocks---without resorting to actual tests on the system. In this
context, state space abstractions are vital when dealing with timed
systems, that exhibit a potentially infinite state spaces. Tina offers
several abstract state space constructions that preserve specific
classes of properties like absence of deadlocks or bisimilarity. A
variety of properties can be checked on abstract state spaces: general
properties---such as reachability properties, deadlock freeness,
liveness, \ldots\,---specific properties relying on the linear
structure of the concrete space state---for example LTL formulas, test
equivalence, \ldots\,---or properties relying on its branching
structure.

Instead of requiring end-users to provide properties written in a
temporal logic, we propose a set of high-level \emph{validation
  patterns} that simplify the elicitation of formal requirements.
This pragmatic approach help us mitigate some of the complexity that
is associated with the use of model-checking tools by novice users.
We have implemented an extension to the frac compiler that accepts the
declaration of realtime specification pattern. Currently, timed
patterns, such as the ``leadsto property'', are compiled into an
observer that is automatically composed with the system at the level
of the Timed Transition System. In the case where the pattern is not
valid, we obtain a counter-example, that is a sequence of events (with
time information) that leads to a problematic state.


\section{Overview of the AADL Translation and Verification of Libraries}

We do not describe precisely the structure of the generated code. In a
nutshell, we associate a pair of Fiacre processes to each AADL thread
and map each AADL port to a communication port in Fiacre. (Since we
focus on the behavior of the system and not its hardware architecture,
we take a flattened view of the AADL model as a set of communicating
threads.)  Timing information, such as the period of threads, are
modeled using the time constraints mechanism provided by Fiacre ports.

The transformation of AADL into Fiacre relies on AADL properties and
on the behavioral annex of AADL that has been developed and integrated
to the OSATE environment. We follow a model-driven approach. Alongside
a meta-model of AADL, we have developed a meta-model of the Fiacre
language that is integrated in the Topcased toolchain. Hence the
transformation from AADL to Fiacre can be obtained through model
transformation.

Our interpretation is fully compositional. Every thread is encoded
using two Fiacre processes, one for its \emph{controller} and another
for encoding its \emph{behavior}. Additional process instances are
created to model the scheduler and the communication resources. The
\emph{controller} process is in charge of the interaction between the
thread and its scheduler (through the ports for the dispatch, complete
and deadline events) and for recovering data from its event data ports
at the right moment. The \emph{controller} process is also in charge
of the ``concurrency protocols'' associated with the shared variables
accessed by the thread (see the discussion at the end of
Section~\ref{sec:aadl-execution-model}). Conversely, the
\emph{behavior} process is used to model the part of the thread
definition associated to its behavior specification (given using the
AADL Behavioral Annex), if any. For the behavior process, the
interpretation of the AADL BA is quite straightforward, since the
behavioral annex is essentially a glorified syntax for a state
transition system. For the controller process, our interpretation
relies on a library of components similar to the code of the
\verb+periodic+ controller given in Listing~\ref{periodic}. We provide
one process for every kind of behavioral resource: threads (periodic,
sporadic, \dots), event and event data port, data connection,
processes and sub-programs (that is schedulers).

The translation takes into account a substantial subset of the AADL
standard and all basic properties are considered when generating a
Fiacre model. More particularly, we take into account AADL priorities,
as well as access to shared variables. For the moment, while periods
can change, we assume that priorities are fixed. Also, we do not take
into account preemption or support for multiprocessor architecture (in
particular we do not take into account the value of the
\verb+Actual_Processor_Binding+ property).

Next, we show how we can use our support for expressing user-defined
properties in Fiacre to check the consistency of our interpretation of
AADL models into Fiacre processes. The correctness of our
interpretation heavily relies on the library of AADL components that
describe the communication and synchronization protocols used to model
the underlying execution model. This library is made of several
\emph{patterns} of Fiacre code that are parameterized by types (the
types of the values exchanged on the communication channels
transferred data); integers (e.g. the size of the communications
queues); and even functions (used for data encoding).  We do not
necessarily know how to check these patterns of Fiacre code
automatically. Therefore, several techniques and tools can be used,
depending on the nature of the component in question: model-checking
can be used in the case of ``finite-state'' code, while theorem
proving techniques may be necessary in the most complex cases. (In a
separate paper~\cite{SCP-aadl1}, some of the authors describe the
framework necessary to carry out proofs on Fiacre specifications using
the Coq assistant prover.) Another source of complexity lies in the
fact that we need to close each code pattern and put it into an
environment that models the context where an AADL component can be
used.

In some cases---like with the controller for AADL periodic
threads---it is possible to generate properties of our embedded
requirement specification language that are enough to prove the
correctness of the code pattern. To avoid the quantification over all
possible context where the code can be inserted, these properties have
to be checked each time a new instance of the AADL component is
created. For example, when checking the correctness of the
interpretation, we need to prove that the system is schedulable;
meaning that the component enters the error state (\verb+sched_error+)
if and only if \verb+c+ (the complete event) is absent between a
\verb+d+ and \verb+dl+ event. This is a consequence of the following
property, (P0a), where \verb+t+ stands for the identifier of the
thread instance.

\begin{lstlisting}[language=FIACRE]
property P0a is ltl [] ((t/event d and 
                       ((not t/event c) until t/event dl)) 
                         => <> t/state sched_error)
\end{lstlisting}

More specifically, in the case of the periodic thread, we also prove
the following list of five requirements. More generally, in our
framework, we provide a specific list of properties for every AADL
component in the library (if it corresponds to a finite state
verification problem).\\

\noindent (P0b) scheduling error implies late completion:
\begin{lstlisting}[language=FIACRE]
property P0b is ltl (([] ((t/event d => 
                     ((not t/event dl) until t/event c)))) 
                    => [] (not (t/state sched_error)))
\end{lstlisting}

\noindent (P1) completion is accepted immediately until scheduling error:
\begin{lstlisting}[language=FIACRE]
property P1 is t/event c leadsto ((t/value (st=p_rdy)) or 
               t/state sched_error) within [0,0] 
\end{lstlisting}
 
\noindent (P2) dispatch is periodic until scheduling error:
\begin{lstlisting}[language=FIACRE]
property P2 is (t/event dl leadsto (t/event dl or 
                t/state sched_error) within [1,1])
\end{lstlisting}

\noindent (P3) deadline is periodic until scheduling error (the event
\verb+t/start+ stands for the initial state of the thread).  We need
to prove property (P3) for every possible period $T$. Nonetheless,
since $T$ is the only timed parameter in this case, it is enough to
consider only one non-null value, say $T=1$.
\begin{lstlisting}[language=FIACRE]
property P3 is (t/event dl or t/start) leadsto 
               (t/event dl or t/state sched_error) within [1,1]
\end{lstlisting}

\noindent (P4) dispatch occurs immediately after deadline:
\begin{lstlisting}[language=FIACRE]
property P4 is t/event dl leadsto t/event d within [0,0]
\end{lstlisting}


Our current toolchain is the result of the refinement and maturation
of several previous versions of the AADL2Fiacre
interpretation~\cite{aadl2fcr,FVAMFT}.  In this most recent iteration
of our tool, we have focused on the modularity of the transformation
with the goal to increase its maintainability and to simplify the
proof of its correctness. Indeed, previous versions of our tool where
based on a monolithic interpretation of AADL, where events and data
exchanges were mediated by a specific glue process that manage
communication and scheduling protocols. Another major contribution of
this work is to define a framework for declaring user-defined
properties at the AADL-level. The same framework is used to generate
``proof-obligations'' that can be checked using our model-checking
toolchain and that ensure that every code pattern is faithful to its
intended semantics.




\section{Experiments}
\label{sec:experiments}

In this section, we report on experiments carried out (1) for
schedulability analysis through model checking and (2) on the dynamic
architecture for a network protocol (NPL) in charge of data
communications between an airplane and ground stations. For the first
study, we observe that model checking allows for a more precise
problem analysis. For the second study, we describe the architecture
of a communication system, the properties that have been checked and
give some quantitative information.

\subsection{Schedulability analysis}

Analytic methods are well known and extensively applied to
schedulability analysis. In order to illustrate their limits, we have
compared the results provided by the Chedar tool\cite{Singhoff:2004} and our
model-checking based tool. The considered example is a typical
non-conservative case which combines dispatch offsets, non preemptive
scheduling and non deterministic execution time. It has been modelled
in AADL but can be summarized by the following table:
\begin{center}
\begin{tabular}{|l|l|l|l|}
\hline
     & Task 1 & Task 2 & Task 3 \\
\hline
period & 20 ms & 20 ms & 20 ms \\
offset & 0        &   3 ms       &  0 \\
deadline & 20 ms & 10 ms & 20 ms \\
priority & 1 (high)       &  2        & 3 (low) \\
BCET..WCET & 1..3 ms & 2 ms & 10 ms \\
\hline
\end{tabular}
\end{center}

As Cheddar does not know how to manage a
BCET..WCET interval, the simulation is done using the WCET bound and
the system is declared to be probably schedulable. However, the
Fiacre-based analysis, through the expression of schedulabilty by
absence of deadlock in some specific state, finds the system
schedulable if the execution time is exactly the WCET and non
schedulable otherwise. Consequently, the Fiacre-based analysis is more
precise. Furthermore, it allows to take into account the
precedence specified by the AADL execution model (linked to immediate
communications). In usual analytic-based schedulability tools like Cheddar, this would require to encode precedence
as priorities and to duplicate threads of which precedence depends on
the task instance. Lastly, the Fiacre-based tool can take into account
data (of finite domain) to make even more fine grain
analysis. However, this method comes at the cost of the model-checking
state exploration.

\subsection{Network protocol}

The considered network protocol, named NPL, implements a communication
protocol based on the Trivial File Transfer Protocol (TFTP) allowing a
pilot and ground stations to receive and send information relative to
the plane: weather, speed, destination, \dots On the hardware side,
the NPL software is running on an IMA computer and consists of one
ARINC 653 partition~\cite{ARINC} that communicates with several other
embedded computers through an AFDX field bus. The dynamic semantics of
these IMA components are taken into account in the AADL model. On the
software side, the protocol layer of the NPL is in charge of handling
messages exchanged between on-board applications and lower ground
systems. Messages are exchanged using a realtime extension of TFTP in
order to ensure predictable response time. For instance, the transport
layer can withstand the loss of messages, which are automatically
re-emitted after a timeout.  Consequently, the NPL stack can be
described using three different layers: a first layer for the
high-level APplications Protocols (APP); the underlying transfer
protocol layer (TFTP in this case); and an intermediate, MiddleWare
Protocol (MWP) layer that mediates the communication between APP and
TFTP.

The overall behavior of the MWP layer can be modeled by a
communicating automaton with three main states (\verb+closed+,
\verb+opening+ and \verb+open+) that correspond to the states of the
``virtual communication'' channel between the aircraft and the ground.
While the number of states is small, the dynamics of the system is
quite complex as it requires about sixty transitions: inputs and
outputs actions of the automaton correspond to requests received or
sent from/to the on-board applications or the lower ground layers. The
complete NPL system is composed of several applications, and every
data-link application has its own instance of the communication
automaton. The main property that should be checked in this context is
the potential accessibility of each state, meaning that the protocol
can always proceed to completion.

\begin{figure}[htbp]
  \begin{center}
    \includegraphics[scale=0.4]{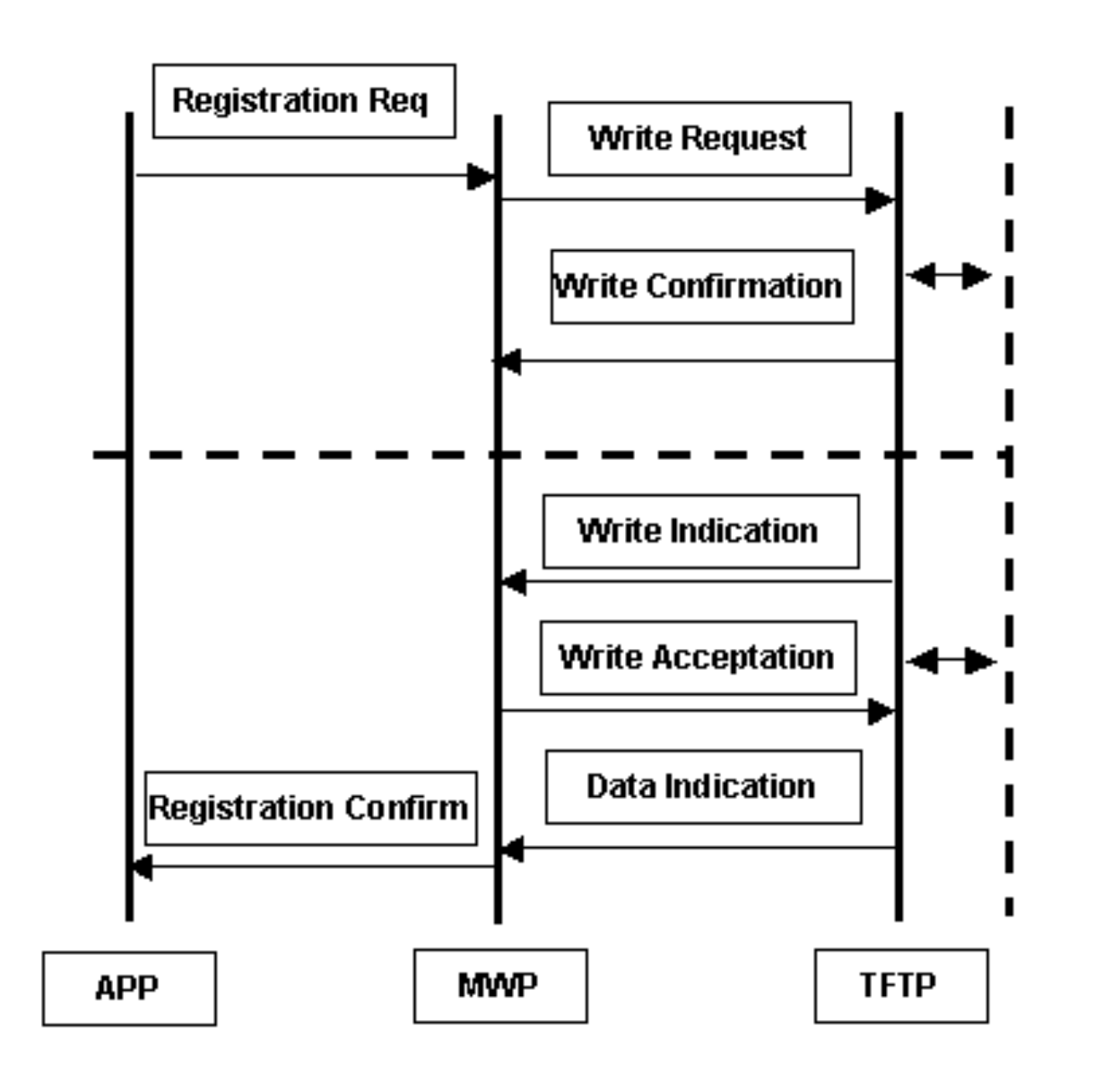}
    \caption{APP registration sequence diagram}
    \label{atc-layer-fig-1}
  \end{center}
\end{figure}

The behavior of the NPL was originally defined by means of sequence
diagrams describing usage scenarios in nominal and default
cases. These sequence diagrams have all been checked against our
automata-based specification in order to assert the correctness of our
modeling. A typical usage scenario is given in
Figure~\ref{atc-layer-fig-1} that details a registration sequence
between an application protocol (APP); the MiddleWare Protocol (MWP);
the transfer protocol; and ground layers tasks (the dashed, vertical
line). This is the most significant activity in the NPL since every
application has to register before starting any data exchanges with
ground stations. The scenarios illustrates two modes of the system. If
the MWP is in state \verb+closed+ and receives a
\verb+registration_request+ message from the APP, it initiates a
connection (the MWP goes into state \verb+opening+). If the MWP is in
the \verb+opening+ state and receives a \verb+data_indication+ message
from TFTP then the connection is established (the MWP enters state
\verb+open+).

\subsubsection{Protocol Modelling with AADL}
\label{sec:prot-modell-with}

\begin{figure*}
\begin{center}
  \includegraphics[width=\textwidth]
  {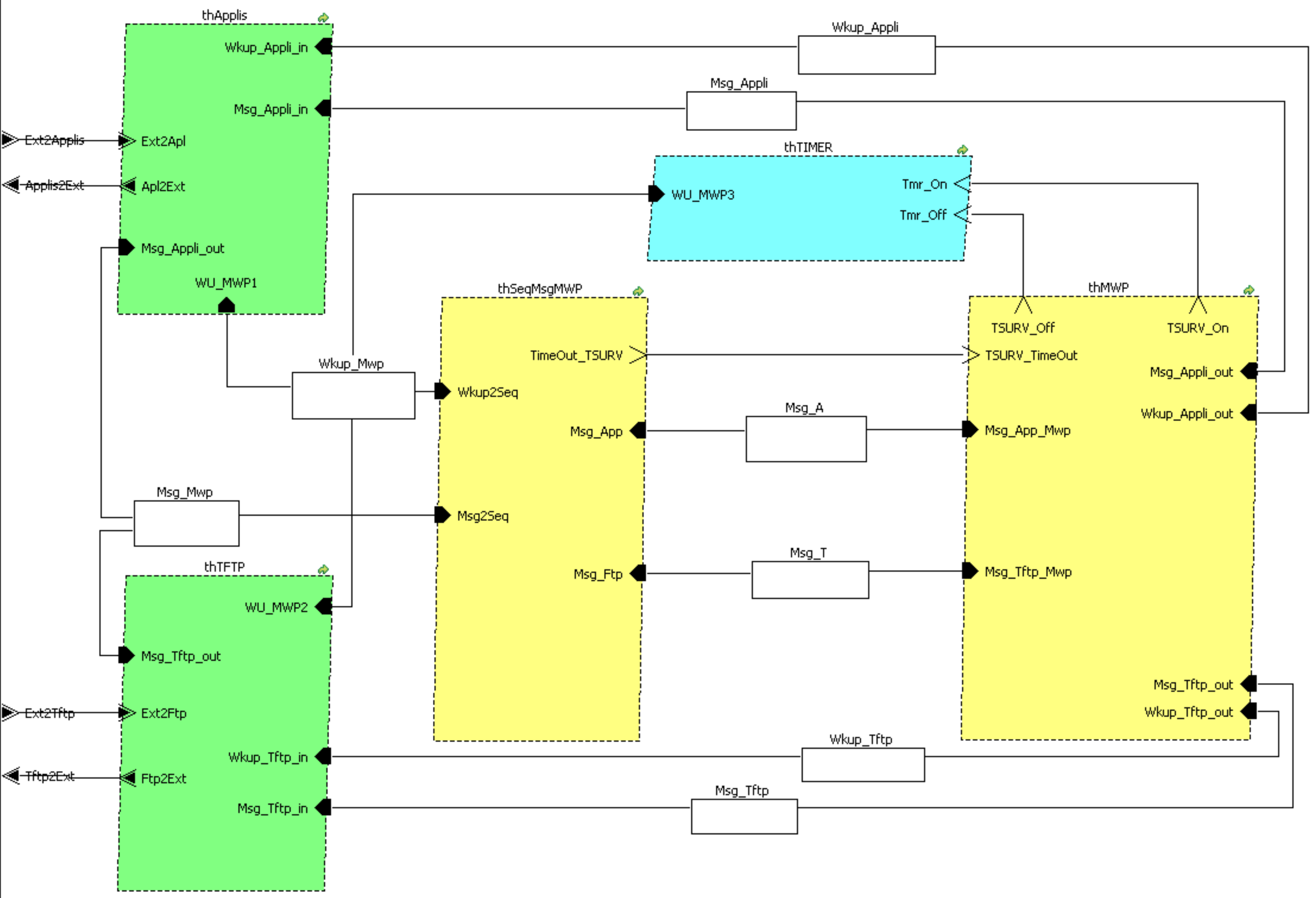}
  \caption{Graphical representation of the NPL component}
\label{graph-atc-fig}
\end{center}
\end{figure*}

The NPL software subset has been modeled as a single application
composed of one main AADL component. This model is mainly derived from
an implementation of the system in the C language provided by Airbus
(see e.g.~\cite{FVAMFT}). The AADL model specifies both the hardware
and software architecture of the component and is composed of: an AADL
processor with its memory (AADL hardware component types); and one
main AADL process that encloses five AADL threads (AADL software
component types). The diagram in Fig.~\ref{graph-atc-fig} details the
architecture of the main AADL process using the AADL graphical
syntax. (This diagram has been edited with the ADELE graphical
modeler~\cite{ADELE}.) We have highlighted the five threads of the NPL
component, which carry out the main functions of the application. A
first thread takes care of the data-link applications
(\verb+thApplis+) while there is another thread for the message
scheduler (\verb+thSeqMsgMWP+). The remaining threads are used for:
implementing the MWP state automaton (\verb+thMWP+); supporting the
Timer functions (\verb+thTIMER+); and supporting the underlying TFTP
protocol (\verb+thTFTP+). In our model, all these threads are periodic
with periods ranging from $5$ ms to $20$ ms.


The architecture of data connections between threads is regular. Each
pair of threads (excluding the timer \verb+thTIMER+) is connected
through at least two memory buffers that are used to store the data
exchanged between the threads over asynchronous communication
channels. A first buffer is used to hold a wake-up signal while the
other carries the message part. For instance, the buffer
\verb+Wkup_Appli+ is used to store the ``wake-up'' signal from the MWP
controller, \verb+thMWP+, to the data-link applications. While this
choice complicates the description of the system, we chose to model
communication between threads using shared data access, instead of
event data port, because it is closer to the actual design found on
avionics software.

All the threads adhere to a common communication protocol. When a
thread needs to communicate with another thread, it first put its
message into the dedicated buffer (for instance \verb+Prim_Appli+) and
then put its identifier into the associated wake-up buffer. When a
thread receives an identifier into its wake-up buffer (pooling), it
reads the message and then clear the identifier.  Some threads have
also access to data generated outside the MWP, like for example
message frames exchanged with the environment, or are connected
through specific event ports (for instance, the timer and the MWP
controller threads). Data exchanged with the environment are defined
as structured, composite data formed from several integer fields.

In our model, the behavior of each thread is expressed using the AADL
Behavioral Annex syntax. The complete AADL specification of the MWP
system requires eight graphical diagrams (of the same complexity than
the one given in Figure~\ref{graph-atc-fig}). In its textual format,
this amounts to about $800$ lines of AADL source code with more than
half of this code automatically generated from the graphical
specification. On these $800$ lines, the behavior of the MWP
controller amounts to about $300$ lines of code. This specification
can be easily reused. Hence, several applications and MWP threads
could be modeled by using several instances of the same AADL
specifications with update connections between them.


\subsubsection{Functional Verification by Model-Checking}
\label{sec:verification}

We used our verification toolchain to check properties on the AADL
specification of the Air Traffic Control system. These properties
corresponds to requirements expressed by the system engineers. Our
experiments were successful as it is possible to verify a substantial
architecture model extracted from the ATC. The main properties
automatically checked on this model can be grouped into three main
categories: (1) \emph{absence of deadlocks}, e.g. the system can not
lock himself due to a wrong synchronization; (2) \emph{healthiness},
e.g. every thread (task) can run and compute infinitely often; and (3)
\emph{absence of dead states}, e.g. every internal behavior state of
every thread is reachable. We also checked several properties related
to the correctness of our interpretation, e.g. to check that the AADL
semantics is preserved in our translation to Fiacre.

The use of formal verification techniques at the model-level is
particularly interesting in the case of the ATC system. Indeed, the
design used in the definition of the communication architecture is
prone to concurrency access problems since all threads must agree on
the same order when accessing data.

We give more details on the properties that have been formally checked
on the model. The goal was to defined a set of simple ``property
patterns'' for dynamic architecture verification and to give them to
avionics software engineers with no previous knowledge of
model-checking or temporal logic. We defined three (untimed) patterns
that were used by system engineers to detect real-time pathologies and
that correspond to the three categories of requirements listed before.
\begin{itemize}
\item \verb+NoGlobalDeadlock+, applies to the whole model. This
  pattern checks for absence of global deadlocks, that is, the system
  can not lock himself due to a wrong synchronization;
\item \verb+Unreachable (exp)+, applies to an internal state. This
  pattern checks for the presence of dead states. This is useful to
  check wether a thread may reach a given behavior state;
\item \verb+Resettable (exp)+, applies to a thread dispatch
  state. This pattern checks for healthiness, that is the fact that a
  given thread can be dispatched infinitely often.
\end{itemize}

These three patterns can be directly encoded in terms of the
LTL-dialect used by the selt model-checker. The pattern
\verb+NoGlobaldeadlock+ is expressed by the formula \verb+[] - dead+,
meaning that for every reachable state (always) it is false that no
transitions can be taken from this state. The pattern
\verb+Unreachable (exp)+ is equivalent to the formula \verb+[]-exp+
(or \verb+absent exp+), meaning that always, the property \verb+exp+
is false. Finally, the pattern \verb+Resettable (exp)+ is equivalent
to the formula \verb+[]<>(exp)+, meaning that always, we will
eventually (after a finite number of transitions) enter in a state
where the property \verb+exp+ is satisfied. For example, the pattern
\verb+Resettable (thApplis/event d)+ can be used to test whether the
thread \verb+thApplis+ will (always) eventually be dispatched.  In
addition to these simple (untimed) patterns, we have also used the
\verb+leadsto+ pattern to find an upper limit on the time needed for
the completion of the sequence diagram given in
Fig.~\ref{atc-layer-fig-1}.

With respect to performances, our verification toolchain is able to
handle the generation of the complete state space of the demonstrator
---which amounts to about 110\,000 states and 150\,000 transitions for
the Fiacre intermediate model---without any memory overflow on a
typical basic development computer (Intel dual-core processor at 2 GHz
clock frequency, and 2 Go of RAM memory). The abstract state space
construction and system compiling are performed, on the same computer,
in less than 5 minutes with a memory footprint in the order of 500 Mo
of RAM. On examples of this size, the model checker included in Tina
is able to generate the whole state space of the system in 15\,s and
to prove a formal properties in a few seconds. For example, it takes
less than 2 minutes to check the 22 properties derived from the
patterns listed before: one test for \verb+NoGlobalDeadlock+; 5
resettable property (one for each thread in the system); and 16
reachability test (one for each state of each thread).

The state space obtained with our new, modular implementation of the
AADL2Fiacre generator is slightly smaller than the one obtained with
our previous, monolithic approach~\cite{FVAMFT}. This is a nice
surprise, since a monolithic interpretation is supposed to produce a
system with less interleaving (and therefore fewer states). The reason
behind this surprising result is that we can use a finer treatment of
priorities between independent threads with a modular approach and
therefore actually reduce the number of interleaving in this case. 

This experimentation, while still modest in size when compared to a
full-blown avionic protocol, gives a good appraisal of the use of
formal verification techniques for real industrial software. These
experimental results are very encouraging. In particular, we can
realistically envisage that system engineers could evaluate different
design choices for the MWP protocol stack in a very short time cycle
and test the safety of their solutions at each iteration.



\section{Related works}

Related work concerning the verification of AADL models is organized
in three subsections: model-checker-based tools for verifying AADL models
through their translation to the input language of existing tools,
model-checking-based verification of the translation to check
intrinsec AADL semantics properties, and analytic methods applying
scheduling analysis techniques to high level abstractions.

\subsection{AADL subsets}
A number of studies have explored how to interpret the AADL standard
in a formal setting. 

A specification of the AADL execution model in
the Temporal Logic of Actions (TLA) is given in~\cite{aadltla5} that
defines one of the earliest formal semantics for AADL. This encoding
takes into account a fixed priority scheduling protocol with
preemption, the management of modes and communication through ports
and shared data. Our approach is based on an interpretation of AADL
specifications, including the Behavioral Annex, in the Fiacre
language. 

A direct encoding from AADL to Petri net is studied
in~\cite{ver06} that takes into account a more limited subset of AADL
(it restricts the behavior of software components and omits realtime
properties of elements). 

Other target formalisms have also been
studied. An encoding of AADL in BIP is presented in~\cite{BIP09P} that
focuses on the behavioral annex as well as on threads, processes and
processors. The approach is improved in~\cite{bipaadlPi} by taking
into account the management of AADL communication protocols. When
compared to BIP, the current version of Fiacre provides less
high-level constructs---therefore encodings are less direct---but
offers better compositional and real-time properties. The library of
AADL component defined in our approach is a first step toward
providing higher-level modeling construct in Fiacre.

Some works consider different technologies for defining the behavior
of software components. In~\cite{Hal07}, the authors study the case
where behaviors are described in a synchronous language, such as Scade
or Lustre. In this case, they define a direct translation that
generate an executable model of the software behavior. Such a model is
usable for early simulation, but also for formal verification, using
tools available for Scade and Lustre. 

Within the COMPASS project, \cite{DBLP:conf/cav/BozzanoCKNNRW10}
  propose the verification of linear or branching time properties on
  AADL-like models with hybrid behaviors. Probabilistic properties are also
  considered. However, the semantics of AADL is not precisely
  considered, while it is one of the main features of our proposal,
  together with the management of time.

The ABV-A verifier\cite{DBLP:conf/iceccs/BjornanderSLP11} does not
translate the AADL model to an existing modelling language. It directly
evaluates temporal logic formulas (written in CTL) on a state space
generated from the AADL model, including the behavior annex. However,
timing information are ignored and the adequacy with the AADL
runtime semantics is not discussed.

In \cite{DBLP:conf/forte/OlveczkyBM10}, the semantics of AADL models
in specified in real-time Maude: timed rewriting rules specify the
update of the system configuration. The time domain can be discrete or
dense. Time advances non deterministically until reaching the date of
the next event. Quantitative linear time properties can be defined and
verified by the Maude model checker. However, the compositional Maude-based
semantics introduces to much asynchrony and leads to inefficient
model-checking. For this purpose a synchronous variant of the tool has
been developped \cite{DBLP:conf/icfem/BaeOAM11}, but takes as input a
subset of AADL (only periodic and synchronous threads, restricted
communication patterns).


Finally, other works~\cite{aadluml6,feileraadl7} have focused on AADL
data communication handling but leave the connection with a formal
verification tool as a perspective.


\subsection{Translation verification}
Another distinctive feature of our work is the concern for checking
the correctness of our interpretation. In this paper, we concentrate
on the definition of ways to check properties on a AADL specification
using model-checking, how to express properties and what kind of
properties can be expressed. In another related work~\cite{SCP-aadl1},
we describe the semantical framework used for the transformation of
AADL into Fiacre and how to check the correctness of this translation
using proof assistant. This companion paper gives more details on the
formal semantics of subsets for both AADL and Fiacre and gives a
high-level description of the translation from one framework to the
other.

\subsection{Analytic methods}

Finally, we have also compared our approach with other AADL related
tools, outside the domain of formal verification, for example with
scheduling analysis tools, such as Cheddar~\cite{cheddar}, that need
to analyze the behavior of (or even simulate) AADL models. Even if
analytical methods outperform model-checker when the scheduling policy
and the analyzed model fall into one of the cases covered by the tool,
we have easyly illustrated the limitations of analytical and
simulation-based approaches using a simple, non-conservative model
combining offsets and non-preemptive scheduling.
 




\section{Conclusion}

This paper describes a formal verification toolchain for AADL that
takes into account the Behavioral Annex. We give a high-level view of
the tools and the transformations involved in our verification
process. While the methodology of our verification toolchain has
already been described in previous works~\cite{aadl2fcr}, this paper
is the first occasion to report on a experimental study that was
conducted on a significant avionic demonstrator. It is also the first
time that we describe our modular interpretation approach as well as
the use of specification pattens to check basic properties on the
correctness of our encoding (such as the schedulability of the
resulting system). This study gives some interesting directions for
further studies. There are several areas for improvements, such as:
enhancing and standardizing our library of AADL component and
validation patterns; improving the behavioral modeling capabilities of
the Adele editor (e.g. with a graphical representation of the
behavioral annex); and improving the integration of the transformation
toolchain in Topcased, in particular with respect to a better
presentation of the verification results to the end user.

Work is still ongoing to improve the tools involved in our
verification framework. A number of extensions to Tina are being
evaluated, concerning new tools, new front-ends, and new
back-ends. For instance, we are experimenting with the addition of
suspension/resumption of actions to Time Petri nets, which is of great
value for modeling scheduled real-time systems. Alongside these works
on tools, our current efforts are directed toward three main
objectives:

(1) \emph{Simplifying the definition of logical properties.} End users
of verification tools should not be required to master temporal
logic. To improve the usability of our approach, we are currently
investigating the proposition of a kit of predefined AADL
requirements or the integration with an AADL-based requirement
specification framework.

(2) \emph{Improving error reporting.} We plan to provide a
``debugging'' procedure, which should take as input a counter-example
produced during the model-checking stage and convert it to a trace
model of the initial AADL description. These traces should be played
back using simulation tools.

(3) \emph{Improving the Verification Process.} We are currently
investigating extensions to the Fiacre language in order to ease the
interpretation of high-level description languages and to optimize the
verification process. One welcome addition would be to integrate the
notion of modes directly in Fiacre. We also plan to address the
problem of specifying scheduling and time-constrained behaviors within
Fiacre. These aspects should have a great impact on the overall
performance of the analysis tool.




\bibliographystyle{elsarticle-num}
\bibliography{ref}

\begin{thebibliography}{}
\expandafter\ifx\csname url\endcsname\relax
  \def\url#1{\texttt{#1}}\fi
\expandafter\ifx\csname urlprefix\endcsname\relax\def\urlprefix{URL }\fi
\expandafter\ifx\csname href\endcsname\relax
  \def\href#1#2{#2} \def\path#1{#1}\fi

\end{thebibliography}


\begin{thebibliography}{10}
\expandafter\ifx\csname url\endcsname\relax
  \def\url#1{\texttt{#1}}\fi
\expandafter\ifx\csname urlprefix\endcsname\relax\def\urlprefix{URL }\fi
\expandafter\ifx\csname href\endcsname\relax
  \def\href#1#2{#2} \def\path#1{#1}\fi

\bibitem{ADELE}
P.~Dissaux, {ADELE}: a versatile system architecture graphical editor based on
  {AADL}, \url{http://gforge.enseeiht.fr/projects/adele/}.

\bibitem{tina}
B.~Berthomieu, P.-O. Ribet, F.~Vernadat, {The tool TINA -- Construction of
  Abstract State Spaces for Time Petri Nets}, Int. Journal of Production
  Research 42(14), (see \url{http://projects.laas.fr/tina/}).

\bibitem{filfmvte2008}
B.~Berthomieu, J.-P. Bodeveix, P.~Farail, M.~Filali, H.~Garavel, P.~Gaufillet,
  F.~Lang, F.~Vernadat, {Fiacre: an Intermediate Language for Model
  Verification in the Topcased Environment}, in: {European Congress on Embedded
  Real-Time Software (ERTS)}, 2008, (see
  \url{http://projects.laas.fr/fiacre/}).

\bibitem{cadp}
H.~Garavel, F.~Lang, R.~Mateescu, W.~Serwe, Cadp 2010: A toolbox for the
  construction and analysis of distributed processes, in: P.~A. Abdulla,
  K.~R.~M. Leino (Eds.), TACAS, Vol. 6605 of LNCS, Springer, 2011, pp.
  372--387.

\bibitem{ttptosfcasd2006}
P.~Farail, P.~Gaufillet, A.~Canals, C.~Le~Camus, D.~Sciamma, P.~Michel,
  X.~Crégut, M.~Pantel, {The TOPCASED project: a Toolkit in Open source for
  Critical Aeronautic SystEms Design}, in: {European Congress on Embedded
  Real-Time Software (ERTS)}, 2006.

\bibitem{aadl2fcr}
B.~Berthomieu, J.-P. Bodeveix, C.~Chaudet, S.~{Dal Zilio}, M.~Filali,
  F.~Vernadat, Formal verification of aadl specifications in the topcased
  environment, in: Ada-Europe 09 -- 14th International Conference on Reliable
  Software Technologies, 2009.

\bibitem{FVAMFT}
B.~Berthomieu, J.-P. Bodeveix, S.~{Dal~Zilio}, P.~Dissaux, M.~Filali, S.~Heim,
  P.~Gaufillet, F.~Vernadat, {Formal Verification of AADL models with Fiacre
  and Tina}, in: ERTSS 2010 -- 5th International Congress and Exhibition on
  Embedded Real-Time Software and Systems, 2010, pp. 1--9.

\bibitem{bipaadl16}
A.~Basu, M.~Bozga, J.~Sifakis, Modeling heterogeneous real-time systems in
  {BIP}, in: SEFM -- IEEE Conference on Software Engineering and Formal
  Methods, 2006.

\bibitem{DalzilioS:fmics2012patterns}
N.~Abid, S.~{Dal Zilio}, D.~{Le Botlan}, {Real-Time Specification Patterns and
  Tools}, in: FMICS 2012 -- 17th International Workshop on Formal Methods for
  Industrial Critical Systems, Vol. 7437 of Lecture Notes in Computer Science,
  Springer-Verlag, 2012, pp. 1--15.

\bibitem{ppsfsv1999}
M.~B. Dwyer, G.~S. Avrunin, J.~C. Corbett, Patterns in property specifications
  for finite-state verification, in: ICSE'99, 1999, pp. 411--420.

\bibitem{HenzingerMP91}
T.~A. Henzinger, Z.~Manna, A.~Pnueli, {Timed Transition Systems}, in: REX
  Workshop, 1991, pp. 226--251.

\bibitem{SCP-aadl1}
J.-P. Bodeveix, M.~Filali, M.~Garnacho, R.~Spadotti, Z.~Yang, {On the
  Mechanization of an AADL Subset}, {Science of Computer Programming : special
  issue on Architecture Design Language}(Submitted).

\bibitem{Singhoff:2004}
F.~Singhoff, J.~Legrand, L.~Nana, L.~Marc{\'e}, Cheddar: A flexible real time
  scheduling framework, Ada Lett. XXIV~(4) (2004) 1--8.

\bibitem{ARINC}
{ARINC} 653 -- avionics application software standard interface, specification
  653, airlines electronic engineering committee (1997).

\bibitem{aadltla5}
J.-F. Rolland, J.-P. Bodeveix, D.~Chemouil, D.~Filali, M.and~Thomas, { Towards
  a formal semantics for AADL execution model}, in: ERTS 2008 -- European
  Congress on Embedded Real-Time Software, 2008.

\bibitem{ver06}
T.~Vergnaud, Mod\'{e}lisation des syst\`{e}mes temps-r\'{e}el r\'{e}partis
  embarqu\'{e}s pour la g\'{e}n\'{e}ration automatique d'applications
  formellement v\'{e}rifi\'{e}es, Ph.D. thesis, \'{E}cole nationale
  sup\'{e}rieure des t\'{e}l\'{e}communications (2006).

\bibitem{BIP09P}
M.~Bozga, V.~Sfyrla, J.~Sifakis, Modeling synchronous systems in {BIP}, in:
  S.~Chakraborty, N.~Halbwachs (Eds.), EMSOFT, ACM, 2009, pp. 77--86.

\bibitem{bipaadlPi}
L.~Pi, J.-P. Bodeveix, M.~Filali, {Modeling AADL Data Communication with BIP},
  in: Ada-Europe 2009 -- 14th International Conference on Reliable Software
  Technologies, 2009, pp. 192--206.

\bibitem{Hal07}
E.~Jahier, N.~Halbwachs, P.~Raymond, X.~Nicollin, D.~Lesens, {Virtual Execution
  of AADL Models via a Translation into Synchronous Programs}, in: EMSOFT --
  ACM \& IEEE international conference on Embedded software, 2007.

\bibitem{DBLP:conf/cav/BozzanoCKNNRW10}
M.~Bozzano, A.~Cimatti, J.-P. Katoen, V.~Y. Nguyen, T.~Noll, M.~Roveri,
  R.~Wimmer, A model checker for aadl, in: T.~Touili, B.~Cook, P.~Jackson
  (Eds.), CAV, Vol. 6174 of Lecture Notes in Computer Science, Springer, 2010,
  pp. 562--565.

\bibitem{DBLP:conf/iceccs/BjornanderSLP11}
S.~Bj{\"o}rnander, C.~C. Seceleanu, K.~Lundqvist, P.~Pettersson, Abv - a
  verifier for the architecture analysis and design language (aadl), in:
  I.~Perseil, K.~Breitman, R.~Sterritt (Eds.), ICECCS, IEEE Computer Society,
  2011, pp. 355--360.

\bibitem{DBLP:conf/forte/OlveczkyBM10}
P.~C. {\"O}lveczky, A.~Boronat, J.~Meseguer, Formal semantics and analysis of
  behavioral aadl models in real-time maude, in: J.~Hatcliff, E.~Zucca (Eds.),
  FMOODS/FORTE, Vol. 6117 of Lecture Notes in Computer Science, Springer, 2010,
  pp. 47--62.

\bibitem{DBLP:conf/icfem/BaeOAM11}
K.~Bae, P.~C. {\"O}lveczky, A.~Al-Nayeem, J.~Meseguer, Synchronous aadl and its
  formal analysis in real-time maude, in: S.~Qin, Z.~Qiu (Eds.), ICFEM, Vol.
  6991 of Lecture Notes in Computer Science, Springer, 2011, pp. 651--667.

\bibitem{aadluml6}
C.~Andr\'{e}, F.~Mallet, R.~{de Simone}, {Modeling of immediate vs. delayed
  data communications: from AADL to UML Marte}, in: Forum on specification \&
  Design Languages, 2007.

\bibitem{feileraadl7}
P.~Feiler, Efficient embedded runtime systems through port communication
  optimization, in: Proc. of ICECCS -- IEEE International Conference on
  Engineering of Complex Computer Systems, 2008.

\bibitem{cheddar}
M.~Kerboeuf, A.~Plantec, F.~Singhoff, A.~Schach, P.~Dissaux, Comparison of six
  ways to extend the scope of cheddar to aadl v2 with osate, in: In proc. of
  {ICECCS}, 2010, pp. 367--372.

\end{thebibliography}

\end{document}